# Hyperbolic-metamaterial waveguides for long-range propagation

Viktoriia E. Babicheva

*Georgia State University, P.O. Box 3965, Atlanta, GA*

*baviev@gmail.com*

**Abstract**. We study optical waveguides that include layers of materials and metamaterials with hyperbolic dispersion (HMM). We consider long-range regime at the dielectric-HMM interface in different waveguide geometries (single interface or symmetric cladding with different layers). In contrast to the traditional analysis of geometrical parameters, we make emphasis on optical properties of constituent materials, and by solving dispersion equations, analyze how dielectric and HMM permittivities affect propagation length and mode size. We derive a figure of merits that should be used for each waveguide in a broad range of permittivity values as well as compare them with plasmonic waveguides. We show conventional plasmonic quality factor, which is the ratio of real to imaginary parts of permittivity, is not applicable to the case of waveguides with complex structure. Both telecommunication wavelengths and mid-infrared spectral ranges are of interest considering recent advances in van der Waals materials such as hexagonal boron nitride. Finally, we evaluate the performance of the waveguides with hexagonal boron nitride in the range where it possesses hyperbolic dispersion (6.2 – 7.3 μm), and we identify the optimum wavelength for each type of the waveguide.

## 1. Introduction

Material and metamaterials with hyperbolic dispersion (hyperbolic metamaterials, HMMs) have attracted avid scientific interest over the last decade because of unusual optical properties and great potential for optical applications [1-4]. These materials support propagating waves with high propagation constant and effective refractive index, and thus allow light concentration at scales much below the diffraction limit [5]. Hyperbolic dispersion enables an increased density of photonic states in the structures that result in strong and broadband spontaneous emission and absorption enhancement as well as anomalous heat transfer and slow light propagation [6-13]. Both natural and artificially designed materials can possess hyperbolic dispersion, but decreasing optical losses in such materials is a long-standing problem and the search for low-losses materials is in progress [14,15].

Recently, HMM were proposed to be included in waveguide components, and in particular to serve as one- or two-side cladding of waveguides [16-25]. HMM-cladded waveguides have been shown to outperform the conventional plasmonic waveguides such as metal-insulator-metal and insulator-metal-insulator waveguides [17]. Furthermore, we have demonstrated theoretically that upon special design, single insulator-HMM (IH) interface can support surface waves with long-range propagation regime [18]. This occurs when perpendicular components of HMM is approximately equal to the permittivity of the adjacent dielectric layer. In contrast to plasmonic waveguides that support long-range waves only being a thin metal film in the nearly uniform cladding [26-30], HMM enable long-range propagation for the waveguide of any thickness, and the propagation length can be on the order of a millimeter [18].

In the present paper, we show the existence of long-range regime for all three waveguides (WGs) – IH, HIH, and IHI – and we show that IHI-WG has the largest propagation-length/mode-size ratio, and HIH-WG has the lowest (Section 2). For each WG, we show how propagation length depends on materials permittivity and what figure of merit (FoM) is the most appropriate for particular permittivity values. We find scaling law of propagation length from real and imaginary parts of permittivity, and we show that these scaling laws are lower than for conventional plasmonic WGs such as IM, IMI, and MIM (Section 3). Among others, we show that neither conventional quality factor $-\text{Re}[\varepsilon]/\text{Im}[\varepsilon]$ nor FoM of propagating surface plasmon polaritons (SPPs) along the single interface $\text{Re}[\varepsilon]^2/\text{Im}[\varepsilon]$ ($\varepsilon$ is metal or HMM permittivity) are applicable to the case of WGs with complex structure. Finally, we analyze optical properties of hexagonal boron nitride in the wavelength range where it possesses type II hyperbolic dispersion. We calculate FoMs at different wavelengths, and we show that for IH-WG, it is beneficial to operate at wavelength $\lambda = 6.9$ μm, and for IHI- and HIH-WGs, at $\lambda \approx 6.2$ μm (Section 4).

## 2. The long-range regime for the WGs with HMM layers

We consider three WG structures that include HMM layers: IH, IHI, and HIH (Fig. 1). The HMM is modeled as an anisotropic medium with effective-permittivity tensor components $(\varepsilon_{xx}, \varepsilon_{yy}, \varepsilon_{zz} = \varepsilon_{xx})$, where $\varepsilon_{yy} > 0$ and $\varepsilon_{xx,zz} < 0$ (type II or metallic-type HMM), and the WGs are considered an infinitely wide (one-dimensional). In all cases, WG claddings are infinitely thick, and we solve dispersion equations for two- or three-layer structure. As dielectric layers, we use materials with permittivity $\varepsilon_c$ close to $\mathrm{Re}[\varepsilon_{yy}]$. As we show below, to achieve long-range propagation, the dielectric permittivity $\varepsilon_c$ should match transverse component of HMM permittivity $\mathrm{Re}[\varepsilon_{yy}]$. In Fig. 1, we plot mode profiles for all three WGs. At the interface, the electric field $E_y$ experiences a discontinuity because $\varepsilon_c = 2$ and $\varepsilon_{yy} = 2.35$; however, in contrast to the conventional metal-dielectric interface, the sign of $E_y$ remains the same. Comparing these three cases for the same $\varepsilon_c = 2$, one can also see that IHI-WG has the highest mode expansion, and HIH-WG has the lowest one.

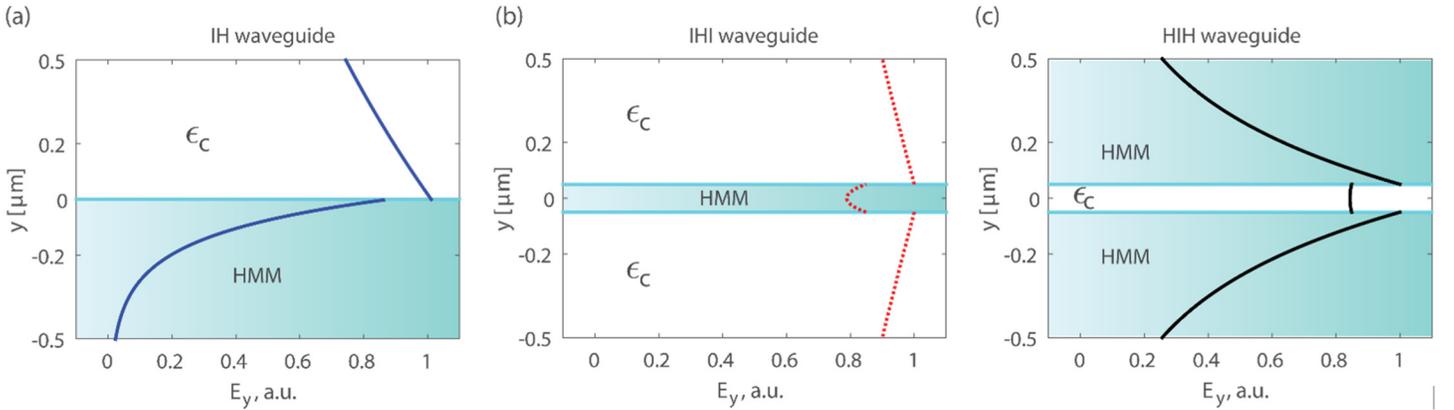

Fig. 1. Schematic view of the WGs under consideration: (a) dielectric-HMM single interface (IH), (b) dielectric-HMM-dielectric waveguide (IHI), and (c) HMM-dielectric-HMM waveguide (HIH). Modes are plotted for a perpendicular component of the electric field $E_y$ and $\varepsilon_c = 2$ for all three WGs. For the IHI-WG, we show only the first symmetric mode. We perform calculations for $d_{\mathrm{IHI}} = d_{\mathrm{HIH}} = 100$ nm, $\varepsilon_{xx} = \varepsilon_{zz} = -24.6 + 0.7i$, and $\varepsilon_{yy} = 2.35 + 0.0002i$, which correspond to effective medium approximation of Ag/MgF$_2$ multilayer with filling ratio $r = 0.2$, $\varepsilon_{\mathrm{Ag}} = -130.35 + 3.5i$ [31], and $\varepsilon_{\mathrm{MgF2}} = 1.878$ [32] for the wavelength $\lambda_0 = 1.55$ μm.

For the single interface between dielectric with permittivity $\varepsilon_c$ and HMM, the mode dispersion is defined by

$$k_{z,\mathrm{IH}} = k_0 \left( \frac{\varepsilon_c (\varepsilon_{xx} - \varepsilon_c)}{\varepsilon_{xx} - \varepsilon_c^2 / \varepsilon_{yy}} \right)^{1/2}. \quad (1)$$

Correspondingly, $k_{z,\mathrm{IHI}}$ and $k_{z,\mathrm{HIH}}$ of IHI- and HIH-WGs can be found from an implicit equation

$$\tanh(k_{y,2}d_2) = -\frac{\left(k_{y,1}/\varepsilon_{zz,1} + k_{y,3}/\varepsilon_{zz,3}\right)k_{y,2}/\varepsilon_{zz,2}}{\left(k_{y,2}/\varepsilon_{zz,2}\right)^2 + k_{y,1}k_{y,3}/(\varepsilon_{zz,1}\varepsilon_{zz,3})}, \quad (2)$$

where for IHI-WG $d_2 = d_{IHI}$ is the HMM thickness, $\varepsilon_{1,zz} = \varepsilon_{3,zz} = \varepsilon_c$, $\varepsilon_{2,zz} = \varepsilon_{zz}$, $k_{y,1} = k_{y,3} = k_{y,\text{diel}} = \sqrt{k_{z,IHI}^2 - \varepsilon_c k_0^2}$, $k_{y,2} = \sqrt{\varepsilon_{xx}(k_{z,IHI}^2/\varepsilon_{yy} - k_0^2)}$; and for HIH-WG $d_2 = d_{HIH}$ is the dielectric thickness, $\varepsilon_{1,zz} = \varepsilon_{3,zz} = \varepsilon_{zz}$, $\varepsilon_{2,zz} = \varepsilon_c$, $k_{y,2} = \sqrt{k_{z,HIH}^2 - \varepsilon_c k_0^2}$, $k_{y,1} = k_{y,3} = k_{y,\text{HMM}} = \sqrt{\varepsilon_{xx}(k_{z,HIH}^2/\varepsilon_{yy} - k_0^2)}$.

Propagation length $L$ of the signal in any WG is defined through a complex propagation constant $k_{z,w}$ of the WG mode as $L = \text{Im}[2k_{z,w}]^{-1}$, and the effective index is $n_{\text{eff}} = \text{Re}[k_{z,w}]/k_0$. Beside it, we study penetration depths of the WG modes that are defined for IH- and IHI-WGs as $\delta_{\text{diel}} = \text{Re}[2k_{y,\text{diel}}]^{-1}$, and for HIH-WG as $\delta_{\text{HMM}} = \text{Re}[2k_{y,\text{HMM}}]^{-1}$. Correspondingly, mode size is $\delta_{\text{IH}} = \delta_{\text{diel,IH}} + \delta_{\text{HMM,IH}}$ for IH-WG, $\delta_{\text{IHI}} = 2\delta_{\text{diel,IHI}} + d_{\text{HMM}}$ for IHI-WG, and $\delta_{\text{HIH}} = d_{\text{HIH}} + 2\delta_{\text{HMM,HIH}}$ for HIH-WG.

To analyze WG properties, we vary $\varepsilon_c$ and fix HMM permittivity $(\varepsilon_{xx}, \varepsilon_{yy}, \varepsilon_{zz} = \varepsilon_{xx})$. Each of IH- and HIH-WGs supports only one mode. However, dispersion equation of IHI-WG has multiple solutions, and we calculate first four modes of it: two symmetric and two anti-symmetric IHI-WG modes. Results of the calculations show that for $\varepsilon_c \approx \text{Re}[\varepsilon_{yy}]$, propagation length in IH-WG, HIH-WG, and IHI-WG with the first mode increases (Fig. 2a). At the same time, no pronounced changes occur for the second, third, and fourth modes of the IHI-WG. One can show that for the long-range modes and relatively low material losses, the inequation $\varepsilon_c < \text{Re}[\varepsilon_{yy}]$ should be satisfied to ensure confinement of the mode to the interface ($\text{Im}[k_{y,\text{diel}}] > 0$ and $\text{Im}[k_{y,\text{HMM}}] > 0$). Similar to the conventional plasmonic WGs, IHI-WG has the highest propagation length and the HIH-WG has the lowest (Fig. 2a).

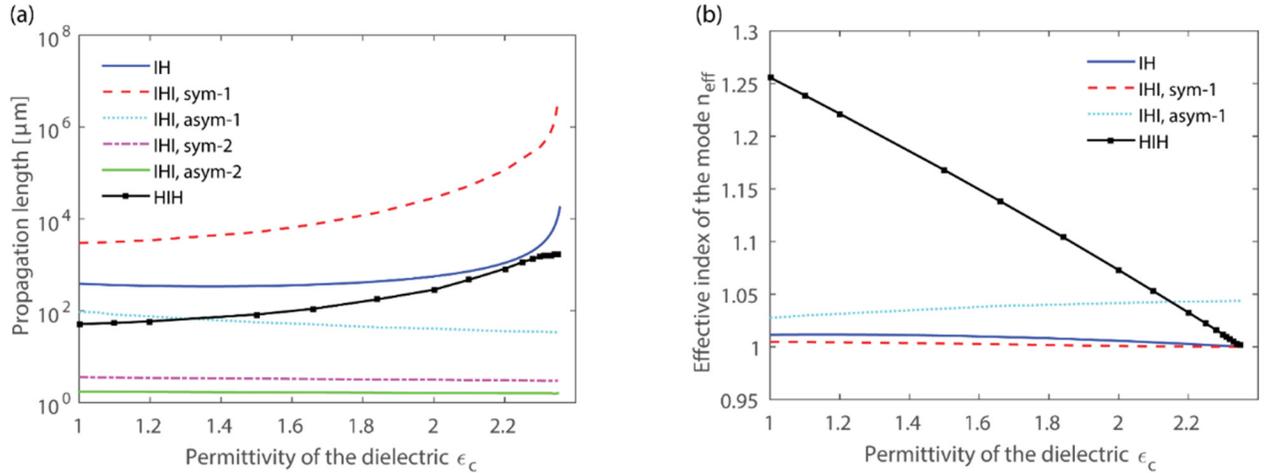

Fig. 2. (a) Propagation length $L$ and (b) effective mode index vs. permittivity of the dielectric $\varepsilon_c$. IH-WG mode, the first symmetric mode of IHI-WG, and HIH-WG mode show a significant increase of $L$ in the vicinity $\varepsilon_c \approx \text{Re}[\varepsilon_{yy}]$. Three other IHI-WG modes monotonically decrease. HMM permittivity and thickness of the WG cores are the same as in Fig. 1.

Effective indices of the modes of IH- and HIH-WGs as well as first two modes of IHI-WG are close to one (Fig. 2b). For the second symmetric and anti-symmetric modes, the indices are 2.1 – 3 and 3.6 – 5.2, respectively (not shown here). HIH-WG possesses a significant decrease of the effective index for $\varepsilon_c \approx \mathrm{Re}[\varepsilon_{yy}]$, which is also typical for the long-range regime.

Furthermore, first symmetric IHI-WG mode as well as IH-WG and HIH-WG modes possess a significant extension in the case of $\varepsilon_c \approx \mathrm{Re}[\varepsilon_{yy}]$ (Fig. 3a). Three other IHI-WG modes (the second symmetric and two anti-symmetric) monotonically decrease. One can also analyze the ratio of the propagation length to effective mode size (Fig. 3b). IH-, IHI-, and HIH-WGs have a common feature: first, the ratio increases for $\varepsilon_c \to \mathrm{Re}[\varepsilon_{yy}]$ and then drops for $\varepsilon_c \approx \mathrm{Re}[\varepsilon_{yy}]$, which means that for $\varepsilon_c \approx \mathrm{Re}[\varepsilon_{yy}]$, the mode size is increasing faster than the propagation length. In contrast, for the second symmetric and anti-symmetric modes, the ratio is approximately 70 and almost does not vary with dielectric permittivity (not shown in the figure).

The ratio of the propagation length to effective mode size is traditionally considered as a figure of merit, but it may be misleading. For instance, results show the ratio is the highest for IHI-WG. However, for the wavelength $\lambda_0 = 1.55$ μm, it has mode size 1-10 μm even out of the long-range regime, which can be impractical for nanophotonic applications. Varying material permittivities (HMM and dielectric) and thickness of the core, one can achieve either higher propagation length or smaller mode size, and the ratio between them will also vary. Thus, the optimum design should be defined based on the application of interest.

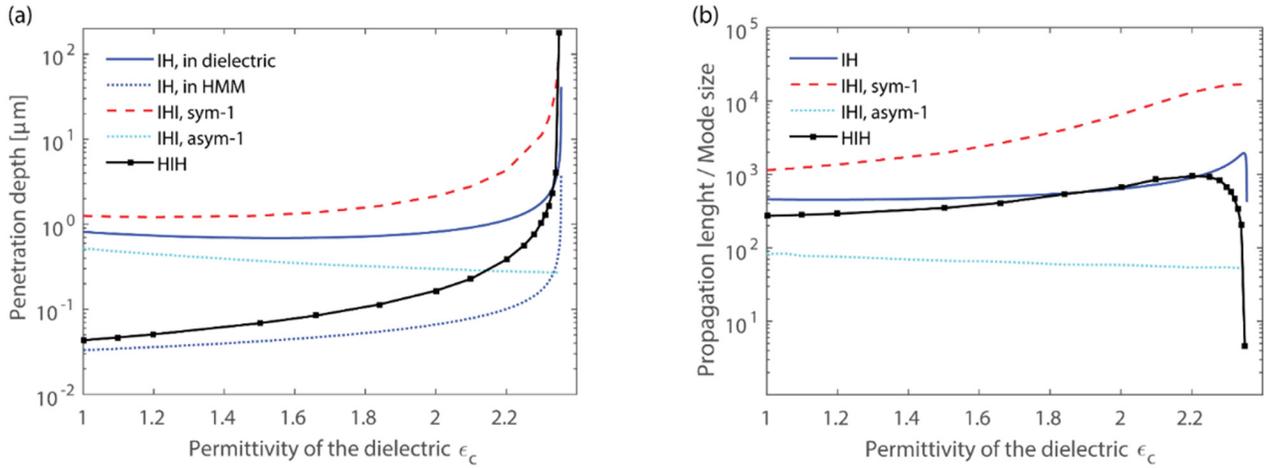

Fig. 3. (a) Penetration depth of the modes outside the IH interfaces. (b) The ratio of the propagation length to effective mode size. HMM permittivity and thickness of WG cores are the same as in Fig. 1.

## 3. Propagation length and figure of merit

For the structures with localized surface plasmon resonance (LSPR) at the metal surface with permittivity $\varepsilon_m$ at frequency ω, the quality factor, or how many plasmon oscillations occur before field decays, $Q_{\mathrm{LSPR}}$ is typically defined as:

$$Q_{\mathrm{LSPR}} = \frac{\omega(\partial \mathrm{Re}[\varepsilon_m(\omega)] / \partial \omega)}{2\,\mathrm{Im}[\varepsilon_m(\omega)]} \qquad (3)$$

or approximately

$$Q_{\mathrm{LSPR}}' = \frac{-\mathrm{Re}[\varepsilon_m]}{\mathrm{Im}[\varepsilon_m]} \qquad (4)$$

Thus, the quality factor of LSPR linearly depends on $\mathrm{Re}[\varepsilon_m]$.

Surface plasmon polaritons on a single insulator-metal (IM) interface have propagation constant $k_{spp} = (\omega/c)\sqrt{\varepsilon_m \varepsilon_c / (\varepsilon_m + \varepsilon_c)}$. One can show that for this waveguide with relatively small losses in the metal, quality factor $Q_{\mathrm{SPP}}$ dependents on $\mathrm{Re}[\varepsilon_m]^2$:

$$Q_{\mathrm{SPP}} = \frac{\mathrm{Re}[k_{spp}]}{2\,\mathrm{Im}[k_{spp}]} \stackrel{\mathrm{Im}[\varepsilon_m] \ll \mathrm{Re}[\varepsilon_m]}{\approx} \frac{\mathrm{Re}[\varepsilon_m]\left(\mathrm{Re}[\varepsilon_m] + \varepsilon_c\right)}{\varepsilon_c \mathrm{Im}[\varepsilon_m]} \stackrel{\varepsilon_c \ll |\varepsilon_m|}{\approx} \frac{\mathrm{Re}[\varepsilon_m]^2}{\varepsilon_c \mathrm{Im}[\varepsilon_m]}. \qquad (5)$$

In turn, propagation length

$$L = \frac{1}{2\,\mathrm{Im}[k_{spp}]} \stackrel{\mathrm{Im}[\varepsilon_m] \ll \mathrm{Re}[\varepsilon_m]}{\approx} \frac{1}{k_0} \frac{\mathrm{Re}[\varepsilon_m]^{1/2} \left(\mathrm{Re}[\varepsilon_m] + \varepsilon_c\right)^{3/2}}{\varepsilon_c^{3/2} \mathrm{Im}[\varepsilon_m]} \stackrel{\varepsilon_c \ll |\varepsilon_m|}{\approx} \frac{1}{k_0} \frac{\mathrm{Re}[\varepsilon_m]^2}{\varepsilon_c^{3/2} \mathrm{Im}[\varepsilon_m]}, \qquad (6)$$

which can also be obtained assuming $\mathrm{Re}[k_{spp}] \approx k_0 \varepsilon_c^{1/2}$ in Eq. (5) and taking into account $\varepsilon_c \ll |\varepsilon_m|$. Thus, we see that FoM of the IM-WG can be defined as

$$FoM_{\mathrm{IM}} = \frac{\mathrm{Re}[\varepsilon_m]^2}{\mathrm{Im}[\varepsilon_m]}. \qquad (7)$$

For more complex WG designs like MIM or IMI, the scaling law is not well defined. Yet, the question is important [33-35]: Recently, a variety of new plasmonic materials has emerged, and one needs to identify the best performing materials. Some materials have low optical losses, but at the same time, $|\mathrm{Re}[\varepsilon_m]|$ is also small [36-39]. Optical properties analyses of different plasmonic materials can be found in Ref. [36] and [37], and here we are focused only on their properties for WG structures.

In Table 1, we summarize metal properties, quality factors $Q_{\mathrm{SPP}}$ and $Q_{\mathrm{LSPR}}'$, and propagation lengths $L_{\mathrm{IM}}$ and $L_{\mathrm{IMI}}$ of IM- and IMI-WGs, respectively. Calculations of $L_{\mathrm{IM}}$ and $L_{\mathrm{IMI}}$ are performed for the WGs with dielectrics of $\varepsilon_c = 2.2$; IMI metal thickness is 20 nm. Different experimental data are available in the literature, including recent advances in the fabrication of low-loss copper and aluminum [40], but we choose data of conventional fabrication techniques. From the calculations, we see that $Q_{\mathrm{SPP}}$ adequately predicts what metal is better for the single-interface IM-WG: $Q_{\mathrm{SPP}}$ and $L_{\mathrm{IM}}$ are the highest for aluminum and the lowest for copper. However, it is not the case for IMI-WG: $L_{\mathrm{IMI}}$ is the highest for gold, and the propagating losses are better described by $Q_{\mathrm{LSPR}}'$.

Table 1. Metals, their permittivities at 1.55 μm, quality factors $Q_{\mathrm{SPP}}$ and $Q_{\mathrm{LSPR}}'$, and propagation lengths $L_{\mathrm{IM}}$ and $L_{\mathrm{IMI}}$ of IM- and IMI-WGs, respectively.

| Metal | $\varepsilon_m$ | $L_{\mathrm{IM}}$ [μm] | $Q_{\mathrm{SPP}}$ | $L_{\mathrm{IMI}}$ [mm] | $Q_{\mathrm{LSPR}}'$ |
|---|---|---|---|---|---|
| Al [41] | -257.5+47i | 109 | 640 | 1.3 | 5.5 |
| Au [42] | -115+11.3i | 87 | 530 | 3.4 | 10.2 |
| Cu [43] | -68+10i | 34 | 210 | 2.4 | 6.8 |

Further, we study power law dependence of propagation length $L$ on real and imaginary parts of permittivity for different WGs under consideration without focusing on specific materials (Fig. 4). For purposes of the study, we vary permittivity in a broad range (somewhere broader than realistic values). Results for this power-law scaling are summarized in Table 2. Calculations for IM-WG agree with Eq. (7) and show that propagation length scales as $\text{Re}[\varepsilon_m]^2$. For IMI- and MIM-WGs, $L$ scales as a lower power of $\text{Re}[\varepsilon_m]$. In particular, for IMI-WG with $-250 \lesssim \text{Re}[\varepsilon_m] \lesssim -25$, the dependence is $L_{\text{IMI}} \sim (-\text{Re}[\varepsilon_m])^{0.58}$. It means that for the case of conventional noble metals, traditional $Q_{\text{SPP}}$ [Eq. (5)] is not applicable at all (see examples in Table 1).

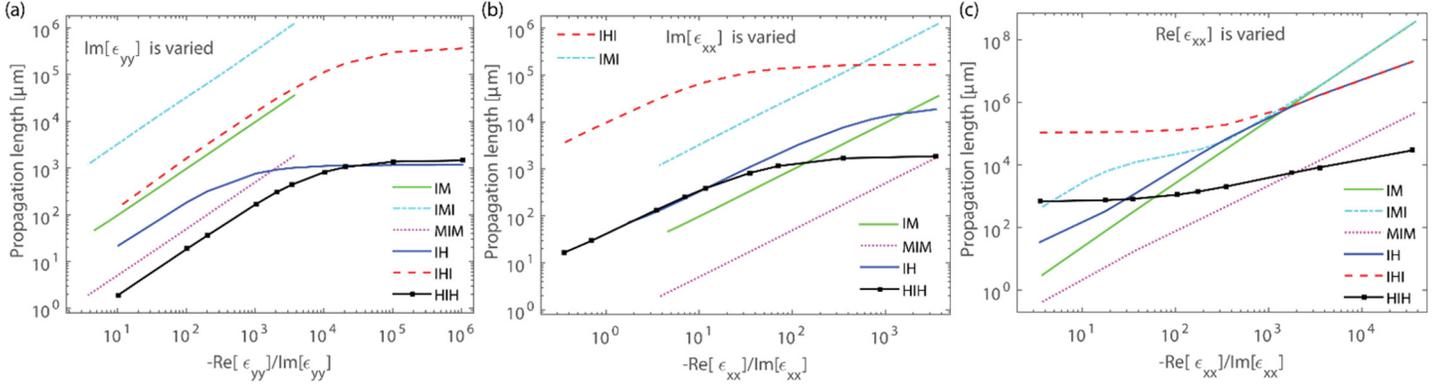

Fig. 4. Propagation length vs. permittivity ratio. (a) $\text{Im}[\varepsilon_{yy}]$ is varied, and $\text{Re}[\varepsilon_{yy}] = 2.35$ and $\varepsilon_{xx} = -24.6+0.7i$ are fixed; (b) $\text{Im}[\varepsilon_{xx}]$ is varied, and $\text{Re}[\varepsilon_{xx}] = -24.6$ and $\varepsilon_{yy} = 2.35+0.0002i$ are fixed; (c) $\text{Re}[\varepsilon_{xx}]$ is varied, and $\text{Im}[\varepsilon_{xx}] = 0.7$ and $\varepsilon_{yy} = 2.35 + 0.0002i$ are fixed. For IM-, IMI-, and MIM-WGs $\varepsilon_{xx} = \varepsilon_{yy} = \varepsilon_{zz}$. IMI-WG has a cladding of $\varepsilon_c = 2.2$ and metal thickness 20 nm, which is thin enough to enable long-range propagation regime for metals. MIM-WG has a gap of 50-nm thickness and $\varepsilon_c = 2.2$. Parameters of IH, IHI, and HIH-WGs are the same as in Figs. 1-3 and $\varepsilon_c = 2.2$ for dielectrics.

Table 2. Power law of real and imaginary parts of permittivity for different WGs: the results of approximation of curves in Fig. 4. For very small losses, some curves become flat, which means that propagation losses are defined solely by the imaginary part of another permittivity component, and thus $k = 0$ ("low-loss" notation).

| WG | $n$ in $L \sim (-\text{Re}[\varepsilon_{xx}])^n$ | | | $k$ in $L \sim (\text{Im}[\varepsilon_{yy}])^{-k}$ or $(\text{Im}[\varepsilon_{xx}])^{-k}$ | |
|---|---|---|---|---|---|
| IM | 2 | | | 1 | |
| IMI | 2 ($\text{Re}[\varepsilon_m] \lesssim -250$) | 0.58 ($-250 \lesssim \text{Re}[\varepsilon_m] \lesssim -25$) | 1.78 ($\text{Re}[\varepsilon_m] \gtrsim -25$) | 1 | |
| MIM | 1.5 | | | 1 | |
| IH | 1 for $(-\text{Re}[\varepsilon_{xx}]) \gtrsim 250$ | 1.5 for $2 \lesssim (-\text{Re}[\varepsilon_{xx}]) \lesssim 250$ | | 1 (regular) | 0 (low losses) |
| IHI | 1 for $(-\text{Re}[\varepsilon_{xx}]) \gtrsim 250$ | 0 for $(-\text{Re}[\varepsilon_{xx}]) \lesssim 250$ | | 1 (regular) | 0 (low losses) |
| HIH | 0.5 for $(-\text{Re}[\varepsilon_{xx}]) \gtrsim 250$ | 0 for $(-\text{Re}[\varepsilon_{xx}]) \lesssim 250$ | | 1 (regular) | 0 (low losses) |

Waveguides of IH, IHI, and HIH configurations in long-range propagation regime possess properties similar to WGs with metals, and propagation length $L$ scales differently in different permittivity ranges (Table 2). As has been shown in the previous subsection, propagation length $L$ strongly depends on the difference between $\text{Re}[\varepsilon_{yy}]$ and $\varepsilon_c$. Thus, a variation of $\text{Re}[\varepsilon_{yy}]$ directly affects long-range regime, and in the further calculations, we keep it fixed. Instead, we vary $\text{Re}[\varepsilon_{xx}]$, $\text{Im}[\varepsilon_{xx}]$, and $\text{Im}[\varepsilon_{yy}]$ with the aim to find $L$ scaling law and define FoMs for different WG containing HMM layers. Two different regimes can be identified based on $\text{Re}[\varepsilon_{xx}]$ value. For in-plane permittivity component with $(-\text{Re}[\varepsilon_{xx}]/\text{Im}[\varepsilon_{xx}]) \lesssim 350$, which means $(-\text{Re}[\varepsilon_{xx}]) \lesssim 250$ for the fixed $\text{Im}[\varepsilon_{xx}] = 0.7$, FoMs of IH-, IHI-, and HIH-WGs in long-range propagation regime ($\text{Re}[\varepsilon_{yy}] \approx \varepsilon_c$) are

$$FoM_{\text{IH}}^{(1)} = \frac{(-\text{Re}[\varepsilon_{xx}])^{3/2}}{\text{Im}[\varepsilon_{xx}]\text{Im}[\varepsilon_{yy}]} \text{ and } FoM_{\text{IHI,HIH}}^{(1)} = \frac{1}{\text{Im}[\varepsilon_{xx}]\text{Im}[\varepsilon_{yy}]}, \qquad (8)$$

and correspondently, for in-plane permittivity component with $(-\text{Re}[\varepsilon_{xx}]) \gtrsim 250$, they are

$$FoM_{\text{IH,IHI}}^{(2)} = \frac{-\text{Re}[\varepsilon_{xx}]}{\text{Im}[\varepsilon_{xx}]\text{Im}[\varepsilon_{yy}]} \text{ and } FoM_{\text{HIH}}^{(2)} = \frac{(-\text{Re}[\varepsilon_{xx}])^{1/2}}{\text{Im}[\varepsilon_{xx}]\text{Im}[\varepsilon_{yy}]}. \qquad (9)$$

**4. Propagating phonon-polaritons in hexagonal boron nitride**

In the previous section, we have derived FoMs for different WG types and considered several metals as an example, but the applicability of these FoMs is not limited to metals and telecommunication wavelength. However, there is a lack of natural materials with hyperbolic dispersion and low optical losses in visible and near-infrared spectral ranges [44], and artificial HMM, e.g. multilayer structures, needs to be included in a WG design. In the case of the multilayer structure, tensor components can be varied by appropriate choice of each layer thickness.

Recently reported studies on van der Waals layers of hexagonal boron nitride have revealed unusual optical properties of the material [45-48]: because of the excitation of phonon-polaritons, permittivity is anisotropic and one or another component can be negative at particular spectral range (two Reststrahlen bands, approximately around $\lambda$ = 7.3 and 12.8 µm). We are interested in the region of $\lambda \approx 6.2$ - 7.3 µm, where hexagonal boron nitride has type II hyperbolic dispersion. In Fig. 5, we show permittivity spectra of data from [49], and only real part is plotted for clarity. Based on it, we calculate FoMs of IH-, IHI-, and HIH-WGs made out of hexagonal boron nitride in different wavelength ranges: $FoM^{(1)}$ is used for $\lambda$ = 6.2 - 6.9 µm, where $(-\text{Re}[\varepsilon_{xx,zz}]/\text{Im}[\varepsilon_{xx,zz}]) \lesssim 350$ and $(-\text{Re}[\varepsilon_{xx,zz}]) \lesssim 10$, and $FoM^{(2)}$ is used for $\lambda$ = 6.9 - 7.3 µm, where $(-\text{Re}[\varepsilon_{xx,zz}]) \gtrsim 10$. Despite Eqs. (8,9) do not take into account change of $\lambda$, the main contribution comes from the change of permittivity, and $\lambda$ does not affect significantly propagation length if long-range propagation regime is achieved at $\text{Re}[\varepsilon_{yy}] \approx \varepsilon_c$. For the spectral range under consideration, $\text{Re}[\varepsilon_{yy}] \approx 2.8$ and does not vary with the wavelength (see Fig. 5a). It means that designing IH-, HIH-, or IHI-WG with long-range propagation, there is no need to adjust dielectric permittivity to a different wavelength and long-range propagation can be achieved with $\varepsilon_c \approx 2.8$. From the FoMs calculations, we see that FoMs vary in a broad range: it is beneficial to operate at $\lambda$ = 6.9 µm designing IH-WG, but $\lambda \approx 6.2$ µm is the best for IHI- and HIH-WGs. For the $\lambda$ > 6.9 µm, the propagation lengths of all WGs drop.

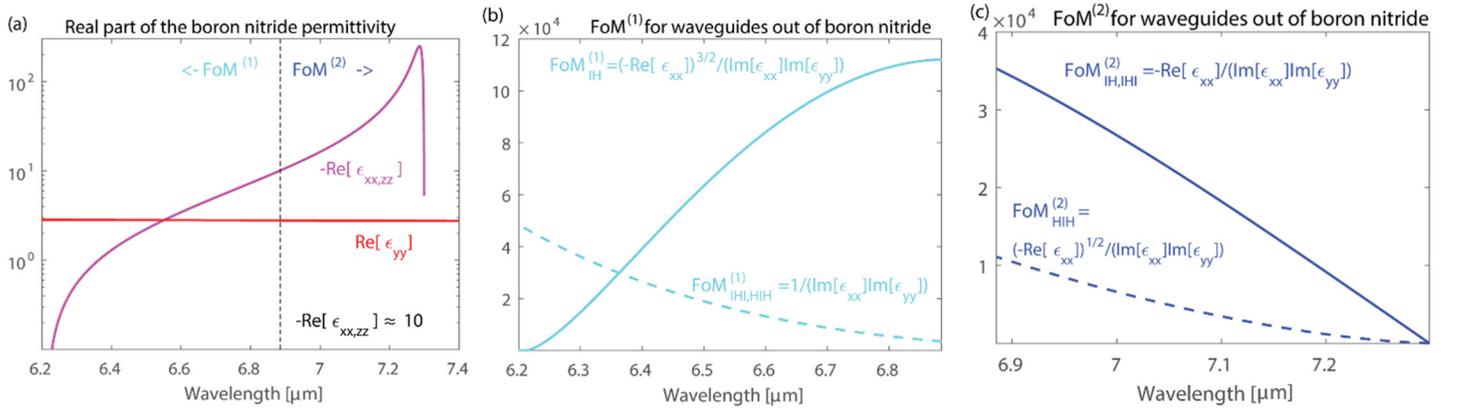

Fig. 5. (a) Tensor components of hexagonal boron nitride permittivity: In-plane $\mathrm{Re}[\varepsilon_{xx}] = \mathrm{Re}[\varepsilon_{zz}]$ and out-of-plane $\mathrm{Re}[\varepsilon_{yy}]$ (only real parts are shown and $\mathrm{Re}[\varepsilon_{xx,zz}]$ is plotted with minus). Black dotted line denotes an approximate change of the regime at $-\mathrm{Re}[\varepsilon_{xx,zz}] = 10$, where different FoMs should be used at larger or smaller wavelengths. (b) $FoM^{(1)}$ for λ = 6.2 - 6.9 μm and (c) $FoM^{(2)}$ for λ = 6.9 - 7.3 μm of IH-, IHI-, and HIH-WGs made of hexagonal boron nitride.

## 5. Conclusion

Long-range propagation of SPPs is well known for thin metal films: it is enabled by effective coupling of plasmons on two metal-dielectric interfaces and can be observed for any permittivity of the dielectric provided that the SPPs exist. In contrast, for the interface of dielectric and HMMs, the possibility of long-range propagation is defined by the dielectric permittivity that needs to match perpendicular component of HMM permittivity tensor. In the present work, we have studied long-range propagation regime for three waveguide designs: single dielectric-HMM interface, dielectric-HMM-dielectric, and HMM-dielectric-HMM. We have shown that similar to a conventional metal-dielectric interface, the dielectric-HMM-dielectric waveguide has the largest propagation length, and HMM-dielectric-HMM waveguide has the best mode localization. We have also studied properties of the three waveguides in long-range propagation regime with respect to material permittivity, both real and imaginary parts. We have compared their dependencies with those of conventional metal waveguides, and in particular, we have shown that propagation length for dielectric-HMM and dielectric-HMM-dielectric is proportional to $\mathrm{Re}[\varepsilon_{xx}]$, while for IM and IMI waveguides it is $\left(\mathrm{Re}[\varepsilon_m]\right)^2$. Finally, we have analyzed properties of naturally-hyperbolic material hexagonal boron nitride and concluded that the best waveguiding properties can be at λ = 6.9 μm for the single interface of dielectric and HMM and at λ ≈ 6.2 μm for the waveguides with symmetric cladding.